\documentclass[12pt]{iopart}
\usepackage{iopams,epsf,psfig}

\def\nothing#1{}

\newcommand{\half}{\mbox{$\textstyle \frac{1}{2}$}}
\newcommand{\quat}{\mbox{$\textstyle \frac{1}{4}$}}

\newcommand{\roothalf}{\mbox{$\textstyle \frac{1}{\sqrt{2}}$}}

\begin{document}
\title[Elementary Derivation for Passage Times]
{Elementary Derivation for Passage Times}

\author[D~C~Brody]{Dorje~C~Brody}
 
\address{Blackett Laboratory, Imperial College, London SW7
2BZ, UK}

\begin{abstract}
When a quantum system undergoes unitary evolution in accordance
with a prescribed Hamiltonian, there is a class of states
$|\psi\rangle$ such that, after the passage of a certain time,
$|\psi\rangle$ is transformed into a state orthogonal to itself.
The shortest time for which this can occur, for a given system, is
called the passage time. We provide an elementary derivation of
the passage time, and demonstrate that the known lower bound, due
to Fleming, is typically attained, except for special cases in
which the energy spectra have particularly simple structures. It
is also shown, using a geodesic argument, that the passage times
for these exceptional cases are necessarily larger than the
Fleming bound. The analysis is extended to passage times for
initially mixed states.
\end{abstract}

\submitto{\JPA}

\section{Introduction}

The notion of a characteristic time arises in a variety of
situations in quantum mechanics. For example, concerning the decay
of an atom, one is interested in the characteristic decay time, or
lifetime. Typically, one would conduct measurements on an ensemble
of independently and identically prepared systems, whereby the
lifetime is estimated as an ensemble mean. For a particle trapped
in a potential, one would be interested in the tunneling time, the
time in which the particle escapes from the trap.

There are many other circumstances in which one is interested in
the time required for an initial state of the system to evolve
into another state under the action of a given Hamiltonian, or
more generally, under some given setup. See, for example,
Ref.~\cite{schulman} (and references cited therein) for a
discussion on various characteristic times in quantum theory. It
is curious that, despite its experimental importance, precise
statistical bounds on the estimation accuracy of time in quantum
mechanics have only been obtained fairly recently
\cite{holevo,brody0}.

One of such characteristic times, namely, the time required for a
given initial state $|\psi\rangle$ to evolve into another state
orthogonal to $|\psi\rangle$, has attracted some attention because
of its relevance to quantum computation and computational capacity
(see, for example, \cite{margolus,lloyd}). Of course, given a
generic state $|\psi\rangle$ and a Hamiltonian, it is more likely
that $|\psi\rangle$ will never evolve into a state orthogonal to
$|\psi\rangle$. Nevertheless, for some special cases this can
occur, which is the situation we study here. In particular, we
call the minimum time required for a state to be transformed into
an orthogonal state  a passage time. The lower bound for the
passage time is known as the Fleming bound \cite{fleming}. Our
main objective here is to give an elementary derivation of the
passage time, and illustrate the result for some simple systems.
Let us first state more explicitly the problem at hand.

Consider an $n$-dimensional Hilbert space ${\mathcal H}$, and a
Hamiltonian ${\hat H}$ with eigenvalues $\{E_l\}$
$(l=1,2,\ldots,n)$. For definiteness, we suppose that the energy
eigenvalues are all distinct, although this is not essential in
the ensuing argument. The time evolution of the wave function is
thus effected by a one parameter family of unitary operators
\begin{eqnarray}
{\hat U}(t) = \exp\left( -{\rm i}{\hat H}t/\hbar\right).
\end{eqnarray}
Now, the Hilbert space ${\mathcal H}$ carries an essentially
redundant complex degree of freedom, i.e. the overall complex
phase associated with the wave function. Thus, we consider
equivalence classes of wave functions, obtained by the
identification
\begin{eqnarray}
|\psi\rangle \sim \lambda |\psi\rangle ,
\end{eqnarray}
where $\lambda\in{\mathbb C}-\{0\}$. In other words, we consider
the space of rays through the origin of ${\mathcal H}$. This is
just the projective Hilbert space ${\mathcal P}$, endowed with the
usual Fubini-Study metric defined by the transition probability
\cite{hughston}. By abuse of notation, we use the symbol
$|\psi\rangle$ to denote both a point of ${\mathcal P}$, {\sl and}
its representative elements in ${\mathcal H}$. This should not
cause confusion.

Given a Hilbert space and a Hamiltonian ${\hat H}$, we seek to
determine the time required for a state $|\psi\rangle$ to be
transformed, under unitary evolution, into another state
$|\eta\rangle$ orthogonal to $|\psi\rangle$. More precisely, the
problem addressed here can be stated as follows:
\begin{itemize}
\item[a)] Does there exist a time $\tau$ such that the state
defined by
\begin{eqnarray}
|\eta\rangle = {\hat U}(\tau)|\psi\rangle
\end{eqnarray}
is orthogonal to $|\psi\rangle$, that is,
$\langle\psi|\eta\rangle=0$, and,
\item[b)] If so, what is the minimum value of $\tau$?
\end{itemize}

Such a minimum time $\tau$, if it exists, will be called the {\sl
passage time}, and denoted by $\tau_{\rm P}$. We shall show that,
in fact, there exist infinitely many, although rather special,
states $|\psi\rangle$ such that $\langle\psi|\eta\rangle=0$ for a
suitable choice of passage time $\tau_{\rm P}$, and that the value
of $\tau_{\rm P}$ for these states is typically given {\sl
exactly} by the expressions
\begin{eqnarray}
\tau_{\rm P} = \frac{\pi\hbar}{\Delta E} = \frac{\pi\hbar}{2\Delta
H}, \label{eq:4}
\end{eqnarray}
where $\Delta E$ and $\Delta H$ are as defined below (note that
the passage time in \cite{schulman} is defined to be given by
$\pi\hbar/2\Delta H$ for an arbitrary state, whereas our
definition here is more refined because we impose orthogonality
condition). There are also cases for which passage times exist but
are larger than $\tau_{\rm P}$ of (\ref{eq:4}). Explicit examples
will be given. We also show, using the Anandan-Aharonov relation,
that (\ref{eq:4}) actually provides the {\sl sharpest obtainable
bound} for the passage time.

\section{Derivation of passage times}

In order to verify (\ref{eq:4}), we first take note of the
Hermitian correspondence between points and hyperplanes of
codimension one in a projective Hilbert space ${\mathcal P}$
\cite{brody}. Specifically, given a point $|\psi\rangle\in
{\mathcal P}$, the corresponding projective hyperplane consists of
those points $|\xi\rangle$ satisfying the algebraic relation
\begin{eqnarray}
\langle\psi|\xi\rangle = 0.
\end{eqnarray}
Thus, if $|\psi\rangle$ is transformed by ${\hat U}(t)$ into a
point $|\eta\rangle$ orthogonal to $|\psi\rangle$, then
$|\eta\rangle$ must lie on this hyperplane, i.e.
$\langle\psi|\eta\rangle=0$. Assuming that such a pair
$(|\psi\rangle, |\eta\rangle)$ of points exists, we can join the
two points by a projective line ${\mathcal P}^1$; the points on
this line represent the totality of normalised superpositions of
the states $|\psi\rangle$ and $|\eta\rangle$. Since a complex
projective line in real terms is just a two-sphere $S^2$, we can
visualise this configuration as illustrated in Figure 1. Note that
the orthogonality of $|\psi\rangle$ and $|\eta\rangle$ implies
that they are {\sl antipodal} on $S^2$. Furthermore, the geodesics
of the Fubini-Study metric that join the two points $|\psi\rangle$
and $|\eta\rangle$ are just the great circle arcs of the sphere
$S^2$ that contain these points.

Next, we observe that, {\it if there exists a unitary evolution
transforming $|\psi\rangle$ into $|\eta\rangle$ along a geodesic
curve, then there must be a pair of energy eigenstates,
$|E_i\rangle$ and $|E_j\rangle$, say, at the poles of $S^2$, such
that $|\psi\rangle$ and $|\eta\rangle$ lie on the equator}. This
is because the dynamics induced by unitary evolution on any
projective line joining a pair of energy eigenstates corresponds
to a rigid rotation of the two-sphere $S^2$ in ${\mathcal P}$,
with the said energy eigenstates as fixed points. Therefore, if we
regard, conversely, the states $|\psi\rangle$ and $|\eta\rangle$
as forming a pair of poles on $S^2$, then the two energy
eigenstates $|E_i\rangle$ and $|E_j\rangle$ will lie on the
corresponding equator. In other words, we have, for some
$\phi\in[0,2\pi)$, the relations
\begin{eqnarray}
\roothalf \left( |\psi\rangle + {\rm e}^{{\rm i}\phi} |\eta\rangle
\right) = |E_i\rangle \label{eq:6}
\end{eqnarray}
and
\begin{eqnarray}
\roothalf \left( |\psi\rangle - {\rm e}^{{\rm i}\phi} |\eta\rangle
\right) = |E_j\rangle , \label{eq:7}
\end{eqnarray}
since $|E_i\rangle$ and $|E_j\rangle$ are antipodal points of
$S^2$. Applying the unitary operator ${\hat U}(\tau)$ to both
sides of (\ref{eq:6}) and (\ref{eq:7}), we obtain
\begin{eqnarray}
\roothalf \left( {\rm e}^{{\rm i}\phi} |\psi\rangle + |\eta\rangle
\right) = {\rm e}^{-{\rm i}E_i\tau/\hbar} |E_i\rangle \label{eq:8}
\end{eqnarray}
and
\begin{eqnarray}
\roothalf \left( -{\rm e}^{{\rm i}\phi}|\psi\rangle +|\eta\rangle
\right) = {\rm e}^{-{\rm i}E_j \tau/\hbar}|E_j\rangle .
\label{eq:9}
\end{eqnarray}
This follows from the fact that, by assumption, the unitary
operator ${\hat U}(\tau)$ for a particular value of $\tau$
interchanges two states $|\psi\rangle$ and $|\eta\rangle$. Thus,
forming the inner products of the respective right and left sides
of (\ref{eq:6}) and (\ref{eq:8}), we find that
\begin{eqnarray}
\half \left( {\rm e}^{{\rm i}\phi} + {\rm e}^{-{\rm i}\phi}
\right) = {\rm e}^{-{\rm i}E_i\tau/\hbar}. \label{eq:10}
\end{eqnarray}
Similarly, from (\ref{eq:7}) and (\ref{eq:9}) we obtain
\begin{eqnarray}
-\half \left( {\rm e}^{{\rm i}\phi} + {\rm e}^{-{\rm i}\phi}
\right) = {\rm e}^{-{\rm i}E_j\tau/\hbar}. \label{eq:11}
\end{eqnarray}
Then, addition of equations (\ref{eq:10}) and (\ref{eq:11}) yields
the condition
\begin{eqnarray}
{\rm e}^{-{\rm i}(E_j-E_i)\tau/\hbar} = -1, \label{eq:12}
\end{eqnarray}
which is satisfied if we set
\begin{eqnarray}
\tau = \frac{\pi\hbar k}{E_j-E_i} \quad (k=1,3,5,\ldots),
\label{eq:13}
\end{eqnarray}
where we assume $E_j>E_i$. Choosing the smallest value for $k$ and
writing $\Delta E=E_j-E_i$ we thus obtain the minimum value
$\tau_{\rm P}$ of the passage time, given by
\begin{eqnarray}
\tau_{\rm P} = \frac{\pi\hbar}{\Delta E} . \label{eq:14}
\end{eqnarray}

\begin{figure}[t]
  {\centerline{\psfig{file=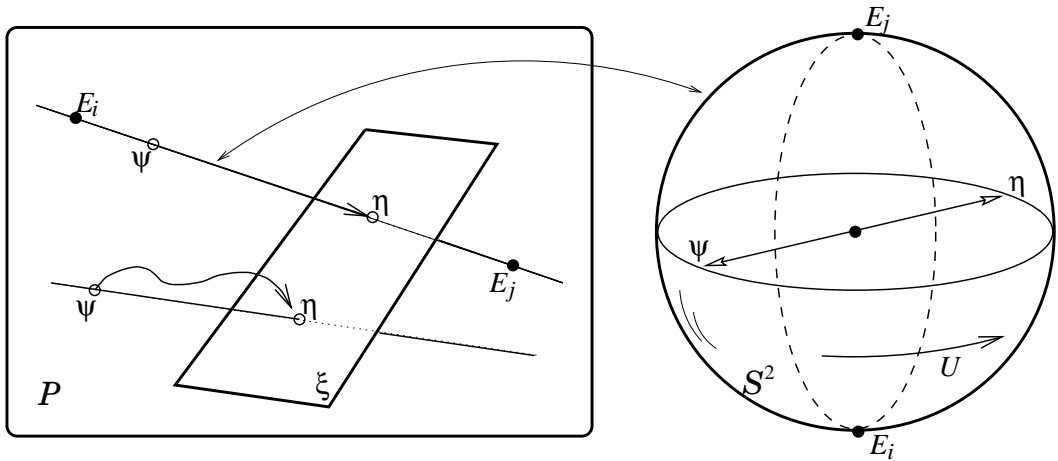,width=14cm,angle=270}}}
   \caption{Hermitian correspondence and projective line. The
   orthogonal complement of a state $|\psi\rangle\in{\mathcal P}$
   is a hyperplane of codimension one such that for any element
   $|\xi\rangle$ on this plane we have $\langle\psi|\xi\rangle=0$.
   If a state $|\psi\rangle$ is transformed into an orthogonal
   state $|\eta\rangle$, then $|\eta\rangle$ must lie on this
   plane. The join of $|\psi\rangle$ and $|\eta\rangle$ is a
   projective line, which in real terms is just a two-sphere
   $S^2$. If the transformation $|\psi\rangle\to|\eta\rangle$
   along a geodesic curve on the projective line is obtained by
   the action of the unitary operators ${\hat U}(t)$, then there
   must be a pair of energy eigenstates $|E_i\rangle$ and
   $|E_j\rangle$ at the poles of the sphere, such that
   $|\psi\rangle$ and $|\eta\rangle$ lie on the equator, and the
   action of ${\hat U}(t)$ is merely a rigid rotation of the
   sphere with respect to these poles. Conversely, if the
   transformations ${\hat U}(t)$ carrying $|\psi\rangle$ into
   $|\eta\rangle$ do not describe a geodesic curve, then there
   exists no pair of energy eigenstates on the projective line
   joining $|\psi\rangle$ and $|\eta\rangle$.
   }
\end{figure}

To summarise, when $|\psi\rangle$ is transformed into an
orthogonal state $|\eta\rangle$ by a one-parameter family of
unitary transformations {\sl along a geodesic curve}, then the
time required is given {\sl exactly} by (\ref{eq:14}). We have not
yet considered the possibility that $|\psi\rangle$ unitarily
evolves into $|\eta\rangle$ along another curve. If an alternative
path exists, then the length of the trajectory is necessarily
longer, since any such path will not be a geodesic. If
$|\psi\rangle$ is expressible as a superposition of $|E_i\rangle$
and $|E_j\rangle$, then the trajectory of ${\hat
U}(t)|\psi\rangle$ never leaves the projective line that joins
these two states, and hence there exists no alternative path. The
case in which $|\psi\rangle$ is expressed as a superposition of
more than two energy eigenstates will be discussed below.

We note, incidentally, that an alternative bound on passage time
was proposed by Margolus and Levitin \cite{margolus}, who {\sl
argued} that a sharper bound for $\tau_{\rm P}$ exists and is
given by the expression
\begin{eqnarray}
\tau_{\rm ML} \geq \frac{\pi\hbar}{2E} , \label{eq:15}
\end{eqnarray}
where $E=\langle{\hat H}\rangle$ is the expectation value of the
Hamiltonian in the state $|\psi\rangle$. However, this inequality
is in general not physically viable, and it is in fact never
sharper than the right-hand side of (\ref{eq:14}). This is because
the physical characteristics of quantum systems are invariant
under an overall shift of the energy spectrum, and hence without
loss of generality we may set, for example, $E=0$ or $E<0$, and
(\ref{eq:15}) becomes meaningless. To avoid this problem, Margolus
and Levitin fix the energy scale so that $E_l\geq0$ for all
$l=1,2,\ldots,n$. Only then does the inequality (\ref{eq:15})
become technically valid. However, this bound, when $2E\geq\Delta
E$, is never attained except in one special case where $E_i=0$, so
that $\Delta E=E_j$ and $2E=E_j$.

\section{Fleming's bound}

We now consider how the passage time $\tau_{\rm P}$ obtained in
(\ref{eq:14}) is related to the dispersion $\Delta H^2 = \langle
({\hat H}-\langle{\hat H}\rangle)^2\rangle$ of the energy. This is
of interest, because a previously derived bound on the passage
time is expressed in terms of the energy dispersion
\cite{fleming}. In the present situation, we can compute $\Delta
H$ explicitly, because the state is expressible in the form
\begin{eqnarray}
|\psi\rangle = \roothalf \left( |E_i\rangle + {\rm e}^{{\rm i}
\varphi} |E_j\rangle \right) \label{eq:17}
\end{eqnarray}
for some $\varphi\in[0,2\pi)$. By a direct calculation, the energy
dispersion in the state (\ref{eq:17}) is
\begin{eqnarray} \Delta
H^2 = \quat\left( E_j-E_i\right)^2, \label{eq:18}
\end{eqnarray}
from which we obtain Fleming's bound
\begin{eqnarray}
\tau_{\rm P} = \frac{\pi\hbar}{2\Delta H} , \label{eq:19}
\end{eqnarray}
as indicated in \cite{schulman}.

This relation is indeed natural if we recall the Anandan-Aharonov
relation \cite{anandan} which states that the `speed' of the
evolution of a given quantum state is given by $2\hbar^{-1}\Delta
H$. The Fubini-Study distance between a pair of orthogonal states
is given by $\pi$, and this distance divided by the velocity
determines the required time. Since the velocity
$2\hbar^{-1}\Delta H$ of the quantum state is a constant under the
action of the unitary group, while the minimum distance of the
trajectory joining a pair of orthogonal states is always $\pi$, it
follows that the Fleming bound can be derived directly from the
Anandan-Aharonov relation.

We have considered thus far the case in which the state
$|\psi\rangle$ is expressible as a superposition of two energy
eigenstates. Next, suppose that $|\psi\rangle$ is expressed as a
superposition of more than two energy eigenstates. It is not
difficult to see that, in this case, if $|\psi\rangle$ can be
transformed into an orthogonal state by a unitary operator ${\hat
U}(t)$, then the energy spectrum $\{E_l\}$ must fulfil rather
stringent constraints. Thus, such a transformation can occur only
for rather special states, in systems such that the energy
spectrum $\{E_j\}$ has a particularly simple structure. In other
words, a generic state in this case will not evolve into an
orthogonal state under the action of ${\hat U} (t)$. It is,
nevertheless, of some interest to analyse such examples in order
to gain further insight into the phenomena involved.

Let us consider, for simplicity, a state $|\psi\rangle$ that is
expressed as a superposition of three energy eigenstates. The most
general form of such a state can be expressed as
\begin{eqnarray}
|\psi\rangle = \cos\alpha |E_i\rangle + \sin\alpha \cos\beta {\rm
e}^{{\rm i}\phi} |E_j\rangle + \sin\alpha \sin\beta {\rm e}^{{\rm
i}\varphi} |E_k\rangle, \label{eq:20}
\end{eqnarray}
where $\alpha,\beta$ are angular coordinates, $\phi,\varphi$ are
phase variables, and we assume that $E_i<E_j<E_k$. If ${\hat
U}(T)$ transforms this state into an orthogonal state, then the
condition
\begin{eqnarray}
\cos^2\alpha + \sin^2\alpha \cos^2\beta {\rm e}^{-{\rm i}
\omega_{ji}T/\hbar} + \sin^2\alpha \sin^2\beta {\rm e}^{-{\rm i}
\omega_{ki}T/\hbar} = 0, \label{eq:21}
\end{eqnarray}
must be satisfied, where $\omega_{ji}=E_j-E_i$ and so on. To
render the analysis more tractible, we further simplify this
constraint by assuming that $\alpha=\beta=\pi/4$. Then,
(\ref{eq:21}) implies that a necessary condition for the state
$|\psi\rangle$ to evolve into an orthogonal state is given by the
relation
\begin{eqnarray}
\frac{\omega_{ki}}{\omega_{ji}} = \frac{2m-1}{2n-1}, \label{eq:22}
\end{eqnarray}
where $m,n$ are natural numbers such that $m\neq n$. Because the
spectrum of a generic Hamiltonian ${\hat H}$ will not satisfy
(\ref{eq:22}), a state $|\psi\rangle$ will never evolve into a
state orthogonal to $|\psi\rangle$. The constraint becomes even
more severe if $|\psi\rangle$ is expressed as a superposition of
more than three eigenstates. The precise form of the constraint in
such cases is just a straightforward generalisation of
(\ref{eq:21}).

Notwithstanding these conditions, let us suppose that the
constraint (\ref{eq:22}) is indeed satisfied for some given
Hamiltonian. Then, the state indeed evolves into an orthogonal
state. The first time that $|\psi\rangle$ becomes orthogonal to
$|\psi\rangle$, in particular, is given by
\begin{eqnarray}
T = \frac{\pi\hbar}{\omega_{ji}} = \frac{3\pi\hbar}{\omega_{ki}} .
\end{eqnarray}
However, {\it since in this case ${\hat U}(t)|\psi\rangle$ does
not describe a geodesic path, $T$ will be larger than Fleming's
passage time $\tau_{\rm P}$ given in} (\ref{eq:19}). Indeed,
without loss of generality, we may set $E_i=0$. Then, it is
straightforward to verify that $T=\sqrt{6}\tau_{\rm P}$. This
follows from the fact that, under the constraint
$\omega_{ki}=3\omega_{ji}$ that follows from (\ref{eq:22}), the
squared energy dispersion in the state (\ref{eq:20}) is given by
$\Delta H^2 = \frac{3}{2}\omega_{ji}^2$.

Another simple example is the cyclic evolution of a spin-1 system,
with energy eigenvalues $-1$, $0$, and $+1$. Consider a state
\begin{eqnarray}
|\psi\rangle = \half |-\rangle + \roothalf |0\rangle + \half
|+\rangle. \label{eq:24}
\end{eqnarray}
The application of ${\hat U}(\pi\hbar)$ yields
\begin{eqnarray}
|\eta\rangle = -\half |-\rangle + \roothalf |0\rangle - \half
|+\rangle,
\end{eqnarray}
and we have $\langle\psi|\eta\rangle=0$. Likewise, the action of
${\hat U}(\pi\hbar)$ on $|\eta\rangle$ yields $|\psi\rangle$,
hence, we have a cyclic evolution that interchanges a pair of
orthogonal states $|\psi\rangle$ and $|\eta\rangle$. However,
because the trajectory ${\hat U}(t)|\psi\rangle$ in ${\mathcal P}$
does not correspond to a geodesic curve, the time required to
interchange these states, given by $T=\pi\hbar$, is longer that
the Fleming bound. Indeed, we have $T=\sqrt{2}\tau_{\rm P}$ in
this example, because in the state (\ref{eq:24}) we have $\langle
H^2\rangle=\frac{1}{2}$ and $\langle H\rangle=0$ so that $\Delta
H^2=\frac{1}{2}$. In general, if a quantum state expressible in
the form other than (\ref{eq:17}) does evolve into an orthogonal
state, then the passage time is necessarily longer than Fleming's
bound (\ref{eq:19}).

\section{Mixed initial states}

The foregoing analysis can be extended in a natural way to the
case in which the initial state of the system is impure. The
situation considered here can be described as follows. Suppose
that we have an initial state, known to be either
$|\psi_1\rangle$, with probability $p$, or $|\psi_2\rangle$, with
probability $1-p$, where both of these pure states are of the form
(\ref{eq:17}). In other words, the initial state is a mixed-state
density matrix
\begin{eqnarray}
{\hat \rho} = p|\psi_1\rangle\langle\psi_1| + (1-p)
|\psi_2\rangle\langle\psi_2| .
\end{eqnarray}
This density matrix evolves in accordance with the Heisenberg law
\begin{eqnarray}
{\hat\rho}(t) = {\hat U}^{\dagger}(t){\hat\rho}{\hat U}(t) .
\end{eqnarray}
Our objective in the present context is to examine the possibility
that, after some lapse of time $\tau_{\rm P}$, the initial pure
state $|\psi_i\rangle$ evolves {\sl with certainty} into a state
orthogonal to $|\psi_i\rangle$, irrespective of whether $i=1$ or
$i=2$.

If the state $|\psi_1\rangle$ is a superposition of energy
eigenstates $|E_i\rangle$ and $|E_j\rangle$, and if
$|\psi_2\rangle$ is a superposition of $|E_k\rangle$ and
$|E_l\rangle$, then the passage time for $|\psi_1\rangle$ is just
$\pi\hbar/\omega_{ji}$, and similarly, for $|\psi_2\rangle$, is
just $\pi\hbar/\omega_{lk}$. Therefore, if the initial state
evolves with certainty into an orthogonal state, then the required
passage time is given by
\begin{eqnarray}
\tau_{\rm P} = \pi\hbar \times {\rm LCM}(\omega_{ji}^{-1},
\omega_{lk}^{-1}), \label{eq:26}
\end{eqnarray}
where ${\rm LCM}(x,y)$ denotes the least common multiple of $x$
and $y$. In other words, since we are uncertain about the initial
state, we must, in general, wait considerably longer before we can
be sure that the state is in another state orthogonal to the
initial state, even though in the meantime the state may evolve
into an orthogonal state and then return to itself many times. It
is straightforward to generalise this argument to the case where
the initial state is one of many states of the form (\ref{eq:17}).
In this case, the passage time is simply given by $\pi\hbar$ times
the least common multiple of the inverses of the energy
differences.

Note that, even though each possible pure state will be
transformed into an orthogonal state after the system has evolved
for the time $\tau_{\rm P}$ given in (\ref{eq:26}), one cannot
clearly argue that the density matrix ${\hat\rho}(\tau_{\rm P})$
has evolved into another mixed state orthogonal to
${\hat\rho}(0)$. Indeed, the diagonal elements of ${\hat\rho}(0)$
and ${\hat\rho}(\tau_{\rm P})$, when expressed in the energy
basis, are identical, and therefore the expectation values of any
observable commuting with the Hamiltonian will also be identical.
This observation leads to an interesting open problem, namely, can
the orthogonality of impure density matrices be defined in a
meaningful fashion, and if so, does a passage time exist for mixed
state density matrices with respect to this definition.

\vspace{0.5cm}
\begin{footnotesize}
\noindent The author acknowledges support from The Royal Society,
and is grateful to L~S~Schulman for posing the problem, as well as
suggesting improvements on an earlier draft of the manuscript.
\end{footnotesize}


\vspace{0.5cm}

\begin{thebibliography}{999}

\bibitem{schulman} Schulman~L~S 2002 ``Jump Time and Passage
Time'' in {\em Time in Quantum Mechanics}, ed. J.~G.~Muga,
R.~Sala~Mayato, and I.~L.~Egusquiza (Berlin: Springer-Verlag)

\bibitem{holevo} Holevo~A~S 1982 {\it Probabilistic and
Statistical Aspects of Quantum Theory} (Amsterdam: North-Holland
Publishing Company)

\bibitem{brody0} Brody~D~C and Hughston~L~P 1996 ``Geometry of
Quantum Statistical Inference'' {\em Phys. Rev. Lett.} {\bf 77}
2581

\bibitem{margolus} Margolus~N and Levitin~L~B 1998 ``The Maximum
Speed of Dynamical Evolution'' {\em Physica} D{\bf 120} 188

\bibitem{lloyd} Lloyd~S 2002 ``Computational Capacity of the
Universe'' {\em Phys. Rev. Lett.} {\bf 88} 237901

\bibitem{fleming} Fleming~G~N 1973 ``A Unitary Bound on the
Evolution of Nonstationary States'' {\em Nuov. Cim.} A{\bf 16} 232

\bibitem{hughston} Hughston~L~P 1995 ``Geometric Aspects of
Quantum Mechanics'' in {\it Twistor Theory}, ed Huggett~S (New
York: Marcel Dekker, Inc.)

\bibitem{brody} Brody~D~C and Hughston~L~P 2001 ``Geometric
Quantum Mechanics'' {\em J. Geom. Phys.} {\bf 38} 19

\bibitem{anandan} Anandan~J and Aharonov~Y 1990 ``Geometry of
Quantum Evolution'' {\em Phys. Rev. Lett.} {\bf 65} 1697


\end{thebibliography}
\end{document}